\documentclass[conference]{IEEEtran}
\IEEEoverridecommandlockouts
\pdfoutput=1 
\usepackage{cite}
\usepackage{amsmath,amssymb,amsfonts}
\usepackage{algorithmic}
\usepackage{algorithm}
\usepackage{graphicx}
\usepackage{textcomp}
\usepackage{makecell}
\usepackage{xcolor}
\usepackage{subfigure}
\usepackage[caption=false,font=normalsize,labelfont=sf,textfont=sf]{subfig}
\usepackage{booktabs}

\newcommand{\hdel}[1]{}

\def\BibTeX{{\rm B\kern-.05em{\sc i\kern-.025em b}\kern-.08em
    T\kern-.1667em\lower.7ex\hbox{E}\kern-.125emX}}
\begin{document}
\title{Transformer-based GAN for Terahertz Spatial-Temporal Channel Modeling and Generating\\
}



\author{
	\IEEEauthorblockN{Zhengdong~Hu, Yuanbo~Li, and Chong~Han}
	\IEEEauthorblockA{Terahertz Wireless Communications (TWC) Laboratory, Shanghai Jiao Tong University, China.\\
	Email: \{huzhengdong, yuanbo.li, chong.han\}@sjtu.edu.cn}
\vspace{-1cm}
}


%
\maketitle

\thispagestyle{empty}
\pdfoptionpdfminorversion=7
\begin{abstract}
Terahertz (THz) communications are envisioned as a promising technology for 6G and beyond wireless systems, providing ultra-broad continuous bandwidth and thus Terabit-per-second (Tbps) data rates. However, as foundation of designing THz communications, channel modeling and characterization are fundamental to scrutinize the potential of the new spectrum. Relied on time-consuming and costly physical measurements, traditional statistical channel modeling methods suffer from the problem of low accuracy with the assumed certain distributions and empirical parameters. In this paper, a transformer-based generative adversarial network modeling method (T-GAN) is proposed in the THz band, which exploits the advantage of GAN in modeling the complex distribution, and the powerful expressive capability of transformer structure. Experimental results reveal that the distribution of channels generated by the proposed T-GAN method shows good agreement with the original channels in terms of the delay spread and angular spread. Moreover, T-GAN achieves good performance in modeling the power delay angular profile, with 2.18 dB root-mean-square error (RMSE).

\end{abstract}

\section{Introduction}
With the exponential growth of the number of interconnected devices, the sixth generation (6G) is expected to achieve intelligent connections of everything, anywhere, anytime~\cite{akyildiz2022terahertz}, which demands Tbit/s wireless 
data rates. To fulfill the demand, Terahertz (THz) communications gain increasing attention as a vital technology of 6G systems, thanks to the ultra-broad bandwidth ranging from tens of GHz to hundreds of GHz~\cite{tera_ref}. 
The THz band is promising to address the spectrum scarcity and capacity limitations of current wireless systems, and realize long-awaited applications, extending from wireless cognition, localization/positioning, to integrated sensing and communication~\cite{han2022thz}.



To design reliable THz wireless systems, one fundamental challenge lies in developing an accurate channel model to portray the propagation phenomena. Due to the high frequencies, new characteristics occur in the THz band, including frequency-selective absorption loss and rough-surface scattering~\cite{yichen_twc}. However, traditional statistical channel modeling methods suffer from the problem of low accuracy with the assumed certain distributions and empirical parameters. For example, a geometric based stochastic channel model (GSCM) assumes that the positions of scatters follow certain statistical distributions, such as the uniform distribution within a circle around the transmitters and receivers~\cite{GSCM}. However, the positions of scatters are hard to characterize by certain statistical distributions, making the GSCM not accurate for utilization in the THz band. To this end, an accurate channel modeling method for the THz band is needed.

Recently, deep learning (DL) is popular and widely applied in wireless communications~\cite{csi}. Among different kinds of DL methods, the generative adversarial network (GAN) has the advantage of modeling complex distribution accurately without any statistical assumptions, based on which GAN can be utilized to develop channel models. The authors in~\cite{appro} train GAN to approximate the probability distribution functions (PDFs) of stochastic channel response. In~\cite{gan-survey}, a GAN based channel modeling method is proposed and demonstrated over a AWGN channel. Nevertheless, these works only address uncomplicated scenarios, necessitating broader applicability to more intricate and practical channels. Considering more complex channels, GAN is designed in~\cite{channel_gan} to generate synthetic channel matrix samples close to the distribution of real channel samples, obtained from clustered delay line (CDL) channel model. In~\cite{distribution}, a model-driven GAN-based channel modeling method is developed in intelligent reflecting surface (IRS) aided communication system. Theses methods~\cite{channel_gan,distribution} learn the distribution of channel matrices to model the channel, and employ convolutional layers to extract image-like features from the channel matrices. However, it is inefficient to directly generate the sparse THz channel matrices with a high dimension, which contain few propagation paths. Moreover, convolutional layers in the GAN are hard to capture the long-range dependencies among elements of channel matrices.

In this paper, a transformer-based GAN spatial-temporal channel modeling method (T-GAN) is proposed in the THz band. In contrast to synthesizing the high-dimensional channel matrices, T-GAN models the channel by generating spatial-temporal channel parameters, which reduces the number of parameters to be learned. Moreover, the transformer structure is integrated to excavate global dependencies among channel parameters. This can enhance the consistency of generated channel parameters, which leads to an improved generation quality of T-GAN. The contributions of this paper are listed as follows.
\begin{itemize}

     \item We formulate the THz channel modeling problem into a task of learning the distribution of spatial-temporal channel parameters. This reduces the required number of learned parameters, compared with learning the high-dimensional channel matrices in the THz band.
    \item We propose a T-GAN based THz spatial-temporal channel modeling and generating method, which integrates the transformer structure with the GAN framework. In this method, T-GAN models the channel by generating channels parameters, including the path gain, phase, delay and azimuth angle of arrival. Relying on the capability of transformer in exploiting the dependencies among the channel parameters, the T-GAN can learn the joint spatial-temporal channel distribution accurately.
    \end{itemize}

The rest of the paper is organized as follows. Sec.~\ref{sec_proposed} details the proposed T-GAN based channel modeling method. Sec.~\ref{sec_performance} demonstrates the performance of the proposed T-GAN method. The paper is concluded in Sec.~\ref{sec_conclusion}.

\textbf{Notation:} 
$a$ is a scalar. \textbf{a} denotes a vector. 
\textbf{A} represents a matrix. $\mathbb{E}\{\cdot \}$ describes the expectation. $\nabla$ denotes the gradient operation. 

\section{Transformer-based GAN Channel Modeling}\label{sec_proposed}
In this section, the channel modeling problem is first formulated into a channel distribution learning problem. Then, the basic framework of the proposed transformer-based GAN (T-GAN) is elaborated. Next, the transformer encoder structure is introduced, which is a key component in the proposed T-GAN. Finally, the detailed structure of T-GAN is presented by integrating the transformer encoder structure into the GAN framework.





\subsection{Problem Formulation}\label{sec_system}
The THz channel can be represented as 
\begin{equation}
    h(\tau) =\sum_{l=0}^{L-1}\alpha_le^{j\phi_l}\delta(\tau-\tau_l),
\end{equation}
which contains $L$ multi-path components (MPCs). Every MPC can be characterized by a set of parameters as 
\begin{equation}
    \mathbf{x}_l = [\alpha_l,\phi_l,\tau_l, \theta_l],
\end{equation}
where $\alpha_l$ denotes the path gain of the $l^{th}$ MPC, $\phi_l$ represents the phase, $\tau_l$ denotes the delay, and $\theta_l$ represents the azimuth angle of arrival (AoA). Then, the THz channel can be characterized by
\begin{equation}\label{eq_channel}
    \mathbf{x} = [\mathbf{x}_1, \mathbf{x}_2,\cdots,\mathbf{x}_{L}],
\end{equation}
where the number of MPCs $L$ is set as 15. The problem of channel modeling can then be described as the generation of channel parameters that forms a distribution of channels. The generating process can be represented by the function 
\begin{equation}
    \hat{\mathbf{x}} = G(\mathbf{z}|c), 
\end{equation}
where $\mathbf{z}$ denotes a random vector sampled from a normal distribution, the variable $c$ is the condition information representing the distance between the transmitter and receiver. Through the function G, the target channel distribution $p_r(\mathbf{x}|c)$ conditioned on the distance can be approximated by the generated distribution $p_g(\hat{\mathbf{x}}|c)$.

\begin{figure}
    \centering
    \includegraphics[width=0.50\textwidth]{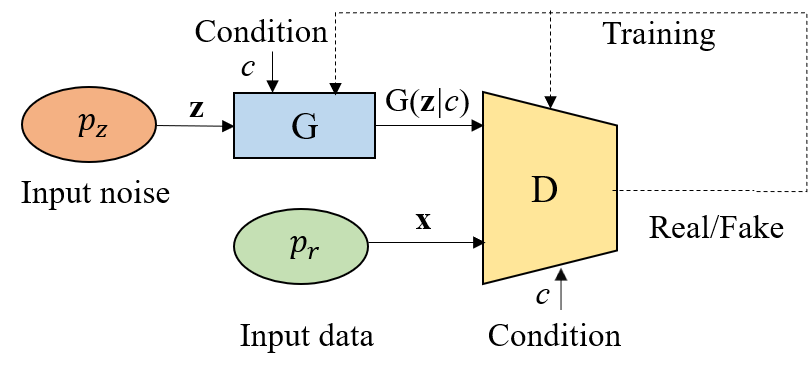}
    \caption{Framework of the proposed T-GAN.}
    \label{fig_gan}
\end{figure}

\subsection{Framework of Proposed T-GAN}
 The T-GAN is designed to learn the channel generating function. The framework of the proposed T-GAN is shown in Fig.~\ref{fig_gan}, which consists of two sub-networks, namely, generator $G$ and discriminator $D$. The generator is aimed at generating the fake channel $G(\mathbf{z}|c)$ conditioned on the distance information $c$ to fool the discriminator, while the discriminator serves as a classifier, trying to distinguish between the real channel $\mathbf{x}$ and fake channel $G(\mathbf{z}|c)$. The two networks are then trained in an adversarial manner, which can be considered as a two-player zero-sum minimax game. Specifically, the training objective can be represented by
\begin{equation}\label{gan_objective}
    \mathop{\min}\limits_{G}\mathop{\max}\limits_{D} \mathbb{E}_{\mathbf{x}\sim p_r}[\log D(\mathbf{x}|c)]+\mathbb{E}_{\mathbf{z}\sim p_z}[\log (1-D(G(\mathbf{z}|c)))],
\end{equation}
where $p_r$ and $p_z$ represent the distributions of real channels and noise vector, respectively. The generator minimizes $(1-D(G(\mathbf{z}|c))$ that represents the probability of the generated channel detected as fake, while the discriminator maximizes this probability. Therefore, the generator and discriminator compete against each other with the opposite objectives in the training process. Through the adversarial training, the Nash equilibrium can be achieved, such that the generator and discriminator cannot improve their objectives by changing only their own network. However, training with the objective function in \eqref{gan_objective} is unstable, since the training objective is potentially not continuous with
respect to the generator’s parameters~\cite{wgan-gp}. Therefore, the improved version of GAN, namely, Wasserstein GAN with gradient penalty~\cite{wgan-gp} is adopted. The modified objective function is expressed as 
\begin{equation}\label{wq_equation}
\begin{aligned}
    \mathop{\min}\limits_{G}\mathop{\max}\limits_{D} \mathbb{E}_{\mathbf{x}\sim p_r}[D(\mathbf{x}|c)]+&\mathbb{E}_{\boldsymbol{z}\sim p_z}[ (1-D(G(\boldsymbol{z}|c)))]\\
      +&\lambda\mathbb{E}_{\tilde{\mathbf{x}}}[(\left\|\nabla_{\tilde{\mathbf{x}}}D(\tilde{\mathbf{x}}|c)\right\|-1)^2)],
\end{aligned}
\end{equation}
where the last term is the gradient penalty term to enforce Lipschitz constraint that the gradient of the GAN network is upper-bounded by a maximum value, the symbol $\tilde{\mathbf{x}}$ is the uniformly sampled point between the points of $\mathbf{x}$ and $G(\mathbf{z}|c)$. Moreover, the parameter $\lambda$ is the penalty coefficient. 
\begin{figure*}[t]
    \centering
    \includegraphics[width=1.0\textwidth]{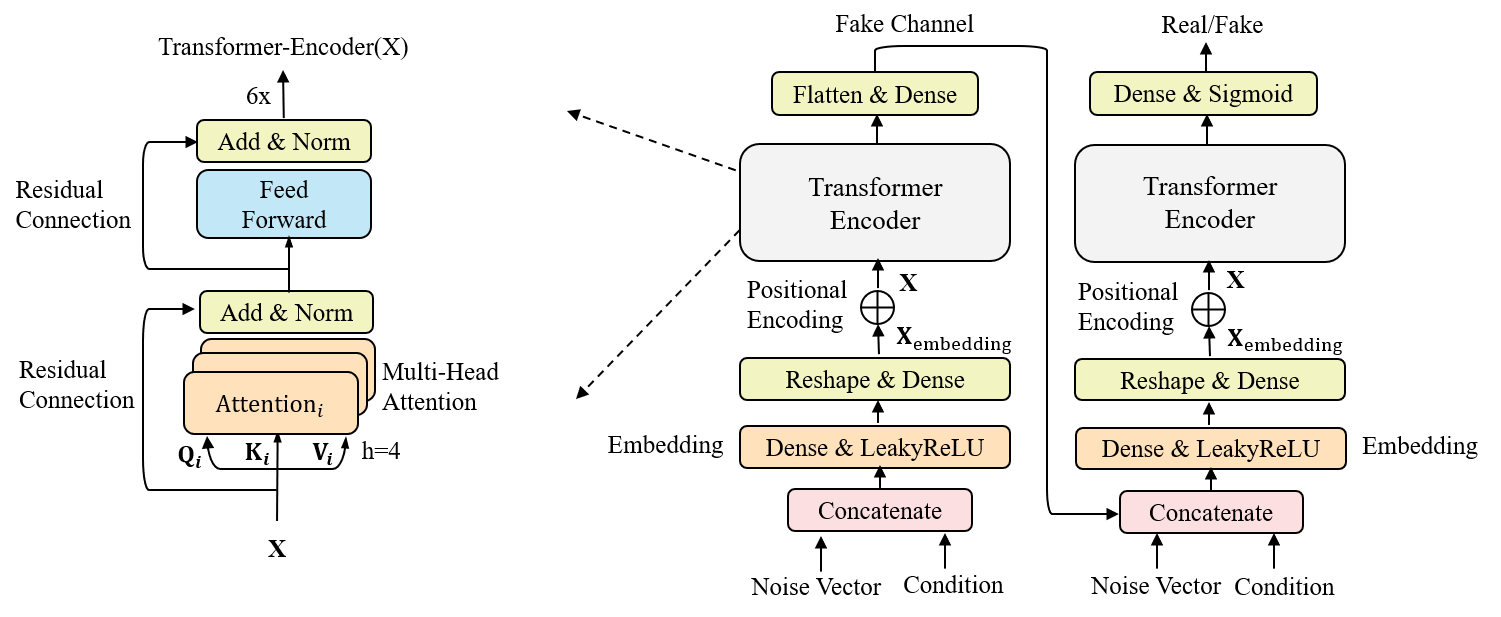}
    \caption{Structure of the transformer encoder (left) and the proposed T-GAN (right).}
    \label{fig_encoder}
\end{figure*}



\subsection{Transformer Encoder Structure}\label{subsec_encoder}
In T-GAN, the channel is inputted as a sequence of MPCs as in~\eqref{eq_channel}. Hence, the transformer enncoder structure can be utilized to capture the dependencies among the MPCs, and the relationships among the parameters in a MPC. As depicted in the left part of Fig.~\ref{fig_encoder}, the transformer encoder consists of 6 stacked identical layers. Every identical layer can be further divided into two sub-layers, multi-head attention layer and feed-forward layer. In both of the two sub-layers, the residual connection is applied by adding the input and the output of the sub-layer represented by $x+\mathrm{Sublayer}(x)$. Moreover, the two sub-layers are followed by layer normalization, which can normalize the input and improve the stability of training.

In the multi-head attention layer, multiple attention layers are applied to the input channel in parallel, so that the model can capture the information of the channel in different subspaces. The implementation of a single attention layer are introduced first. Considering a input channel $\mathbf{X}=(\mathbf{x}_1,\cdots,\mathbf{x}_L)\in \mathbb{R}^{L\times d_x }$, it is composed of $L$ MPCs and every MPC is represented by a vector $\mathbf{x}_l\in \mathbb{R}^{1\times d_x}$. Firstly, every MPC in the sequence is transformed by
\begin{align}
    \mathbf{q}_l &= \mathbf{x}_l\mathbf{W}^q,\\
    \mathbf{k}_l &= \mathbf{x}_l\mathbf{W}^k,\\
    \mathbf{v}_l &= \mathbf{x}_l\mathbf{W}^v,
\end{align}
where $\mathbf{W}^q\in \mathbb{R}^{d_x\times d_k}$, $\mathbf{W}^k\in \mathbb{R}^{d_x\times d_k}$, $\mathbf{W}^v\in \mathbb{R}^{d_x\times d_v}$ are the learned transformation parameters. The symbols $\mathbf{q}_l\in \mathbb{R}^{1\times d_k} $, $\mathbf{k}_l\in \mathbb{R}^{1\times d_k} $ and $\mathbf{v}_l\in \mathbb{R}^{1\times d_v}$ denote query, key and value respectively. The correlation between the query vector and the key vector shows how much attention should be paid to the value vector in the output. To give a concise representation, the vectors are packed into matrices represented by
\begin{align}
    \mathbf{Q} &= \mathbf{X}\mathbf{W}^q,\\
    \mathbf{K} &= \mathbf{X}\mathbf{W}^k,\\
    \mathbf{V} & = \mathbf{X}\mathbf{W}^v,
\end{align}
where $\mathbf{Q}\in \mathbb{R}^{L\times d_k }$, $\mathbf{K}\in \mathbb{R}^{L\times d_k}$, $\mathbf{V}\in \mathbb{R}^{L\times d_v}$ are the matrix representations of query, key and value. Then, the output can be calculated as
\begin{equation}
    \mathrm{Attention}(\mathbf{Q},\mathbf{K}, \mathbf{V}) =\mathrm{softmax}(\frac{\mathbf{QK^T}}{\sqrt{d_k}})\mathbf{V},
\end{equation}
where $\mathrm{Attention}(\mathbf{Q},\mathbf{K}, \mathbf{V})\in \mathbb{R}^{L\times d_v}$ is the output of the attention layer, the term $\mathrm{softmax}(\frac{\mathbf{QK^T}}{\sqrt{d_k}})$ is the calculated attention matrix assigned to the value vector in matrix $\mathbf{V}$. The $\mathrm{softmax}$ is the operation for normalizing the attention weights, defined as
\begin{equation}
    \mathrm{softmax}(\mathbf{x}) =  \frac{e^{x_i}}{\sum e^{x_i}},
\end{equation}
where $x_i$ is the element in vector $\mathbf{x}$, and the $\mathrm{softmax}$ operation ensures that the sum of the output equals one. 

With the single attention layer introduced, the multi-head attention layer is formed by concatenating the result of $h=4$ attention layers, which can be represented by
\begin{align}
    \mathrm{Head}_i &= \mathrm{Attention}(\mathbf{Q}_i,\mathbf{K}_i, \mathbf{V}_i),\\
    \mathbf{X}^o  & = \mathrm{Concat}(  \mathrm{Head}_1,   \mathrm{Head}_2,\cdots, \mathrm{Head}_h)\mathbf{W}^o,
\end{align}
where $i = 1,\cdots,4$ indexes the attention layer, the term $\mathrm{Head}_i\in\mathbb{R}^{L\times d_v}$ denotes the result of the $i^{th}$ parallel attention layer, $\mathbf{W}^o\in \mathbb{R}^{hd_v\times d_x}$ is the linear matrix that transforms the concatenated result $\mathbb{R}^{L\times hd_v}$ into the output $\mathbf{X}^o\in\mathbb{R}^{L\times d_x}$.


The output of the multi-head attention layer is then passed to the feedforward layer, which is just two dense layers with ReLU activation. The ReLU activation function is defined as
\begin{equation}
    f(x) = \mathrm{max}(0,x).
\end{equation}
Then, the feedforward operation can be characterized by
\begin{equation}
    \mathrm{FFN(\mathbf{X}^o)} = \mathrm{max}(0,\mathbf{X}^o\mathbf{W_1}+\mathbf{b}_1)\mathbf{W}_2+\mathbf{b}_2,
\end{equation}
where $\mathbf{X}^o\in \mathbb{R}^{L\times d_x}$ denotes the input to the feedforward layer. Moreover, $\mathbf{W}_1\in \mathbb{R}^{d_x\times d_x}$ and $\mathbf{W}_1\in \mathbb{R}^{d_x\times d_x}$ are the linear transformation matrices, and $\mathbf{b}_1\in \mathbb{R}^{d_x\times 1} $ and $\mathbf{b}_2\in\mathbb{R}^{d_x\times 1}$ are the bias terms for the two dense layers.

\subsection{Structure of Proposed T-GAN}
The detailed architecture of proposed transformer based GAN network is shown in the right part of Fig.~\ref{fig_encoder}. The input to the generator includes the noise vector $\mathbf{z}\in \mathbb{R}^{32\times 1}$ and the condition variable $c\in\mathbb{R}^{1\times 1}$. In the Embedding layer, the two inputs $\mathbf{z}$ and $c$ are first concatenated into $\mathbb{R}^{33\times 1}$, and are then transformed by one dense layer with LeakyReLU function into vector $\mathbb{R}^{Ld_x\times 1}$, where $L=15$ and $d_m=4$. The LeakyReLU function is represented by
\begin{equation}
    f(x) = \begin{cases}
	      x, & \mathrm{if} \ x \geq0 \\
	      \alpha x, & \mathrm{if} \ x <0	
		   \end{cases},
\end{equation}
where $\alpha$ is the slope coefficient when the value of neuron $x$ is negative. Then, the vector is reshaped into the matrix $\mathbb{R}^{L\times d_m}$, and are linearly transformed into the sequence $\mathbf{X}_{\mathrm{embedding}}\in \mathbb{R}^{L\times d_{x}}$ with one dense layer. The parameter $d_x=128$ is the dimension of the embedding representation. The Embedding layer is then followed by the positional encoding, to encode the position information into the sequence $\mathbf{X}$. The operation can be represented by
\begin{align}
    \mathbf{X} &= \mathbf{X}_{\mathrm{embedding}} + \mathbf{PE},
\end{align}
where $\mathbf{PE}\in \mathbb{R}^{L\times d_{x}}$ is the learned positional information of the sequence $\mathbf{X}$. Furthermore, the encoded sequence is forwarded to the transformer encoder structure as introduced in Sec.~\ref{subsec_encoder}. Following the transformer structure, one Flatten layer and two dense layers are applied to get the output of generator $\hat{\mathbf{x}}\in\mathbb{R}^{60\times 1}$. The two dense layers have 240 and 60 neurons, respectively. Then, together with the condition variable $c$, the fake channel $\hat{\mathbf{x}}$ or real channel $\mathbf{x}\in \mathbb{R}^{60\times 1}$ is passed to the discriminator. 

The structures of the discriminator and generator are symmetric, with the similar embedding and transformer encoder structure, except that the noise vector in the generator is replaced by the real channel or fake channel in the discriminator. In the Embedding layer, the channel and condition variable are concatenated and transformed. Then, the position encoding learns the position information. Afterwards, the transformer encoder structure are applied. Next, the output of the transformer structure is transformed by two dense layers both with only one neurons. Finally, the Sigmoid activation function restricts the output of the discriminator in the range of [0,1], which is defined by
\begin{equation}
    f(x) = \frac{1}{1+e^{-x}}.
\end{equation}

\begin{figure}[t]
    \centering
    \includegraphics[width=0.5\textwidth]{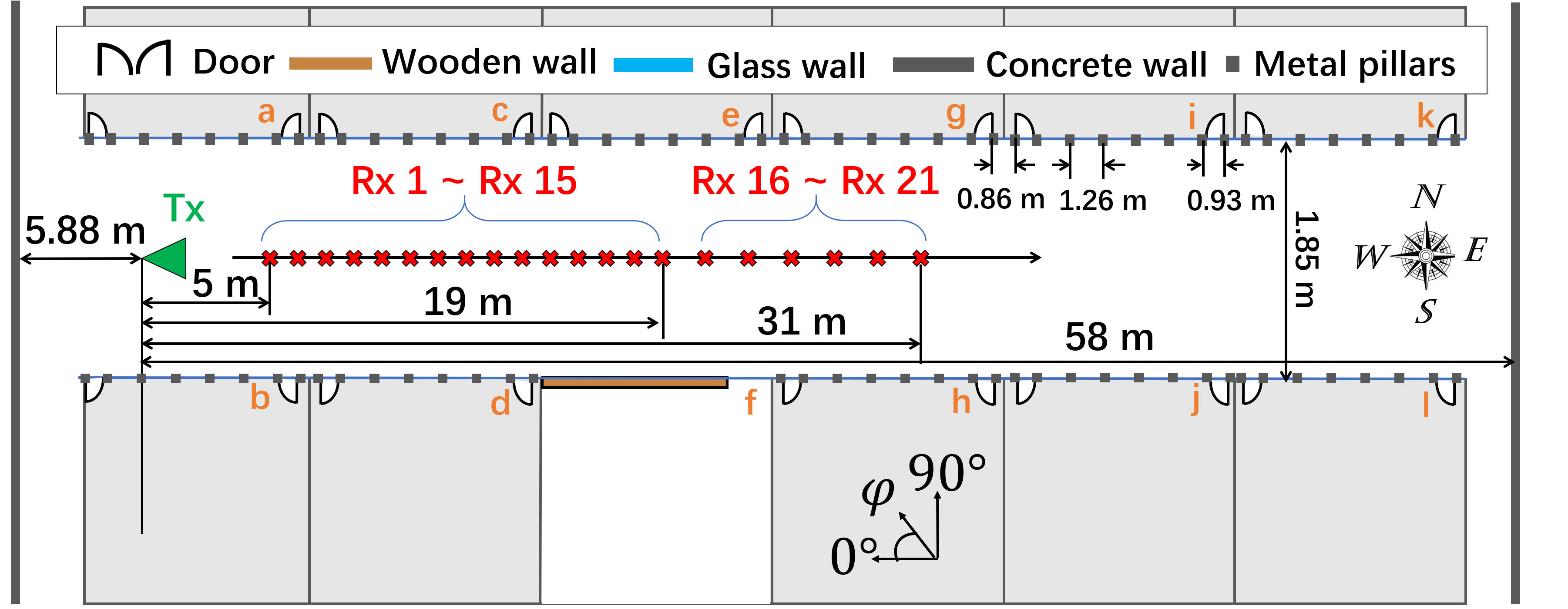}
    \caption{Measurement layout in the indoor corridor scenario~\cite{yuanbo_icc}.}
    \label{fig_scenario}
\end{figure}

\section{Experiment and Performance Evaluation}\label{sec_performance}
In this section, the experiment settings including the dataset and training procedures are elaborated. Moreover, the performance of the T-GAN are evaluated by comparing distributions of the generated channels with the modelled original channels, in terms of delay spread, angular spread and power delay angular profile.

\subsection{Dataset and Setup}
 In the experiment, the dataset is generated by QuaDRiGa~\cite{quadriga} with the extracted statistics from the THz measurement~\cite{yuanbo_icc}. The measurement campaign is conducted in an indoor corridor scenario at 306-321 GHz, as depicted in Fig.~\ref{fig_scenario}. The dataset consists of 50000 channel samples, in which 80$\%$ of the dataset is for training and 20 $\%$ is for testing. Moreover, every channel sample can be represented by a number of $L=15$ MPCs as in~\eqref{eq_channel}, and every feature of MPCs including path gain, phase, delay and AoA angle is normalized into the range of [0,1] by min-max standardization. Besides, the angle of the line-of-sight (LoS) is set at zero degree, which provides a reference point for the generating of other MPCs. 





The training procedure of the proposed GAN network is explained in detail as follows. Firstly, the input noise vector $\mathbf{z}\in\mathbb{R}^{32\times 1}$ is generated by the multivariate normal distribution, which can provide the capabilities to transform into the desired distribution. The gradient penalty parameter $\lambda$ in \eqref{wq_equation} is set as 10, which works well in the training process. Moreover, the stochastic gradient descent (SGD) optimizer is applied for the generator network, and the adaptive moment estimation (Adam) optimizer is chosen for the discriminator network. In addition, the learning rates of the two optimizers are both set as 0.0001 to stabilize the training. The number of epochs for training the proposed T-GAN is set as 10000. A epoch is defined as a complete training cycle through the training dataset, during which the generator and discriminator are trained iteratively, once and three times, respectively.




All the experimental results are implemented on a PC with AMD Ryzen Threadripper 3990X @ 2.19 GHz and one Nvidia GeForce RTX 3090 Ti GPUs. In addition, the training of GAN network is carried out in the Tensorflow framework.

\begin{figure*}[h]
    \centering
    \subfigure[Delay spread.]{\includegraphics[width=0.43\textwidth]{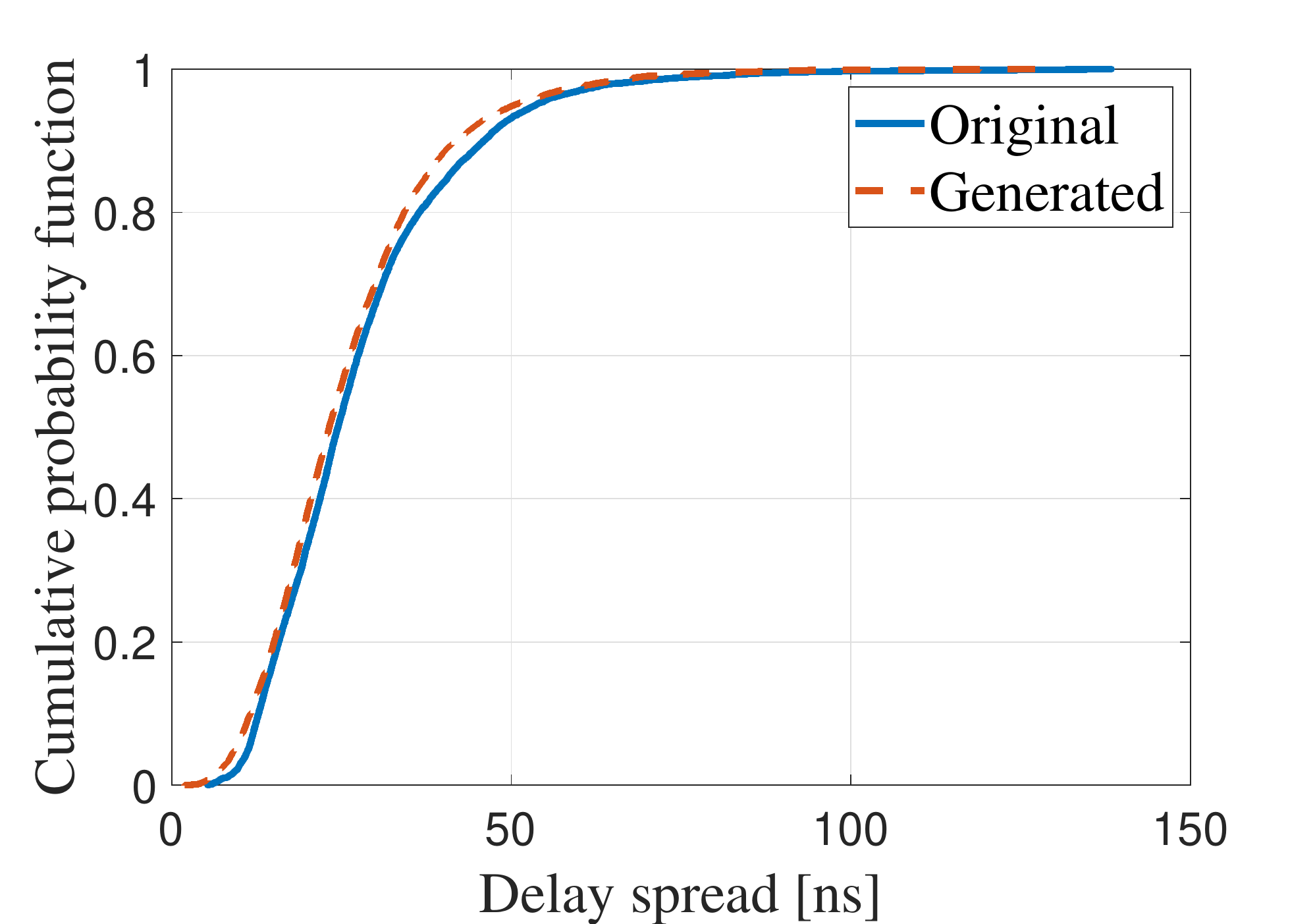}}
    \subfigure[Delay spread.]{\includegraphics[width=0.43\textwidth]{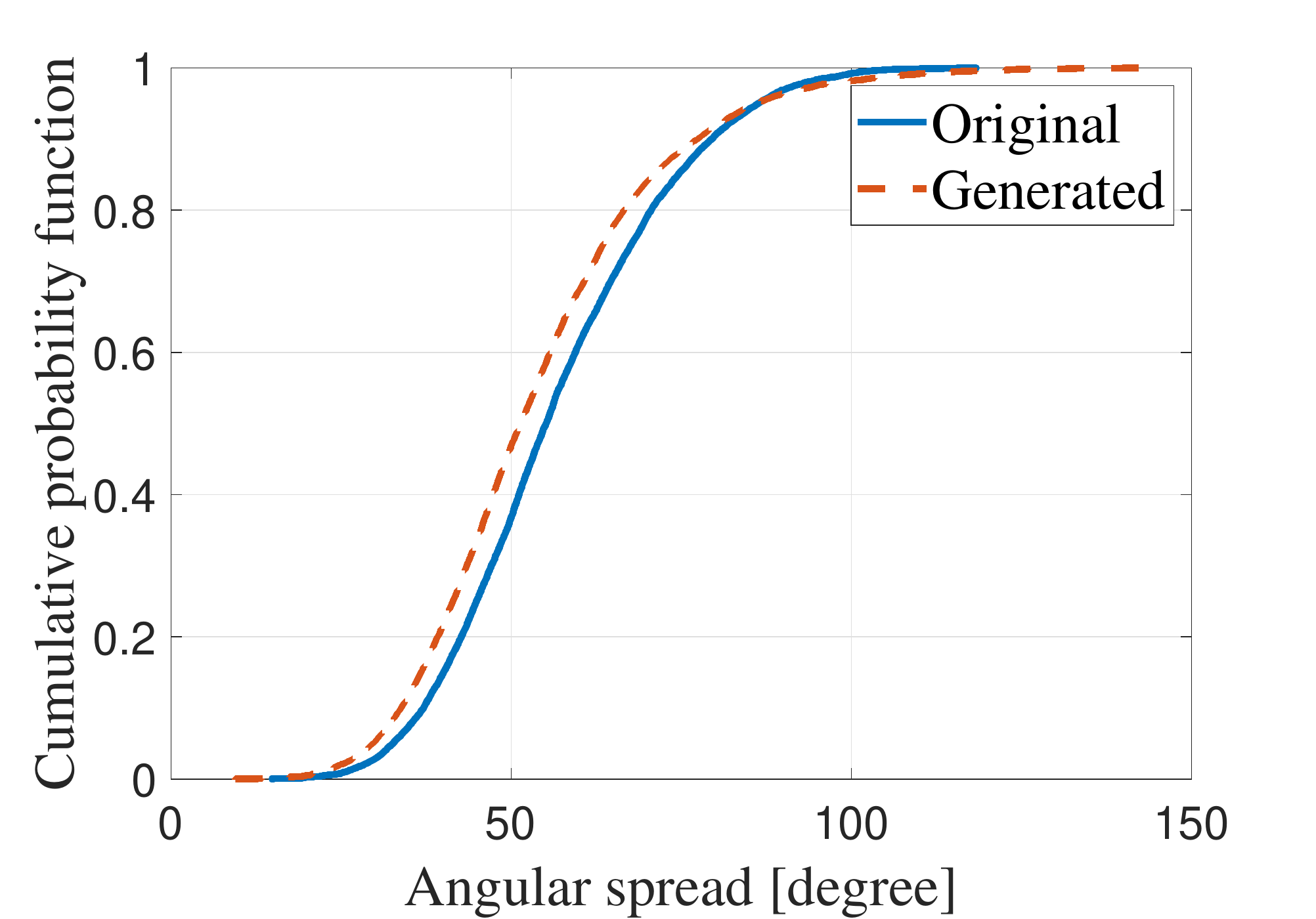}}
    \caption{Delay spread and angular spread for the original and generated channels.}
     \label{fig_delay}
    \end{figure*}

\begin{figure*}
\centering
     \includegraphics[width=1.0\textwidth]{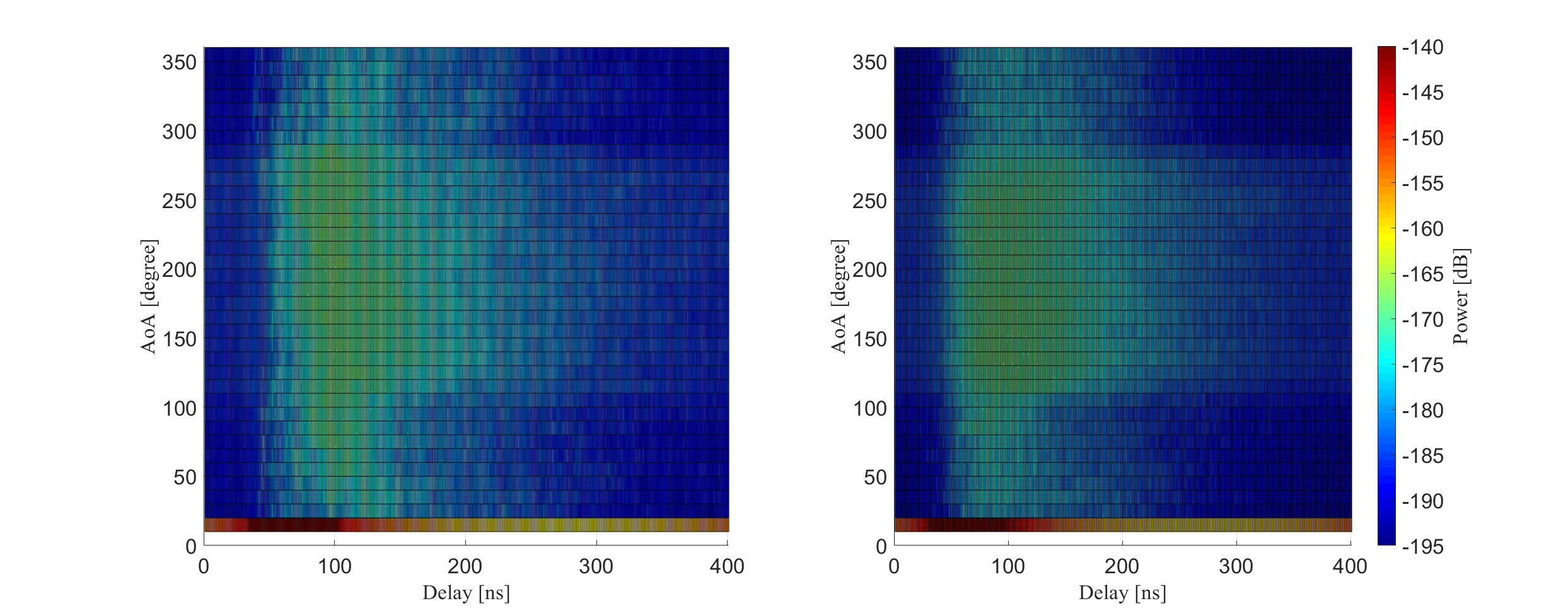}
    \caption{Average PDAP for the original (left) and generated channels (right). }
    \label{fig_pdap}
\end{figure*}

\subsection{Delay Spread}
Delay spread characterizes the power dispersion of multi-path components in the temporal domain. It is an important metric to measure the small-scale fading, which can be computed by
\begin{equation}
\begin{split}
\Bar{\tau}&=\frac{\sum_{i=0}^{N_{\tau}}i\Delta \tau P_{\tau}(i)}{\sum_{i=0}^{N_{\tau}}P_{\tau}(i)},\\
    \tau_{rms}& = \sqrt{\frac{\sum_{i=0}^{N_{\tau}}(i\Delta \tau-\Bar{\tau})^2P_{\tau}(i)}{\sum_{i=0}^{N_{\tau}}P_{\tau}(i)}},
\end{split}
\end{equation}
where $N_{\tau}$ denotes the number of sampling points in the temporal domain, $\Bar{\tau}$ denotes the mean delay weighted by the power, $\tau_{rms}$ refers to the root-mean-square (RMS) delay spread, $\Delta \tau$ denotes the sampling time interval, and $P_{\tau}(i)$ denotes the power at the delay of $i\Delta\tau$. The cumulative probability function (CDF) plot of delay spread for the original and generated channels is depicted in Fig.~\ref{fig_delay}(a). It can be observed that the CDF of delay spread for the generated channels matches the original channels well. Moreover, the average values of delay spread for the original and generated channels are 27.42 ns and 25.67 ns, respectively, which are very close. This shows that the T-GAN can well capture the channel characteristics in the temporal domain.  


\begin{figure}
    \centering
    \includegraphics[width=0.43\textwidth]{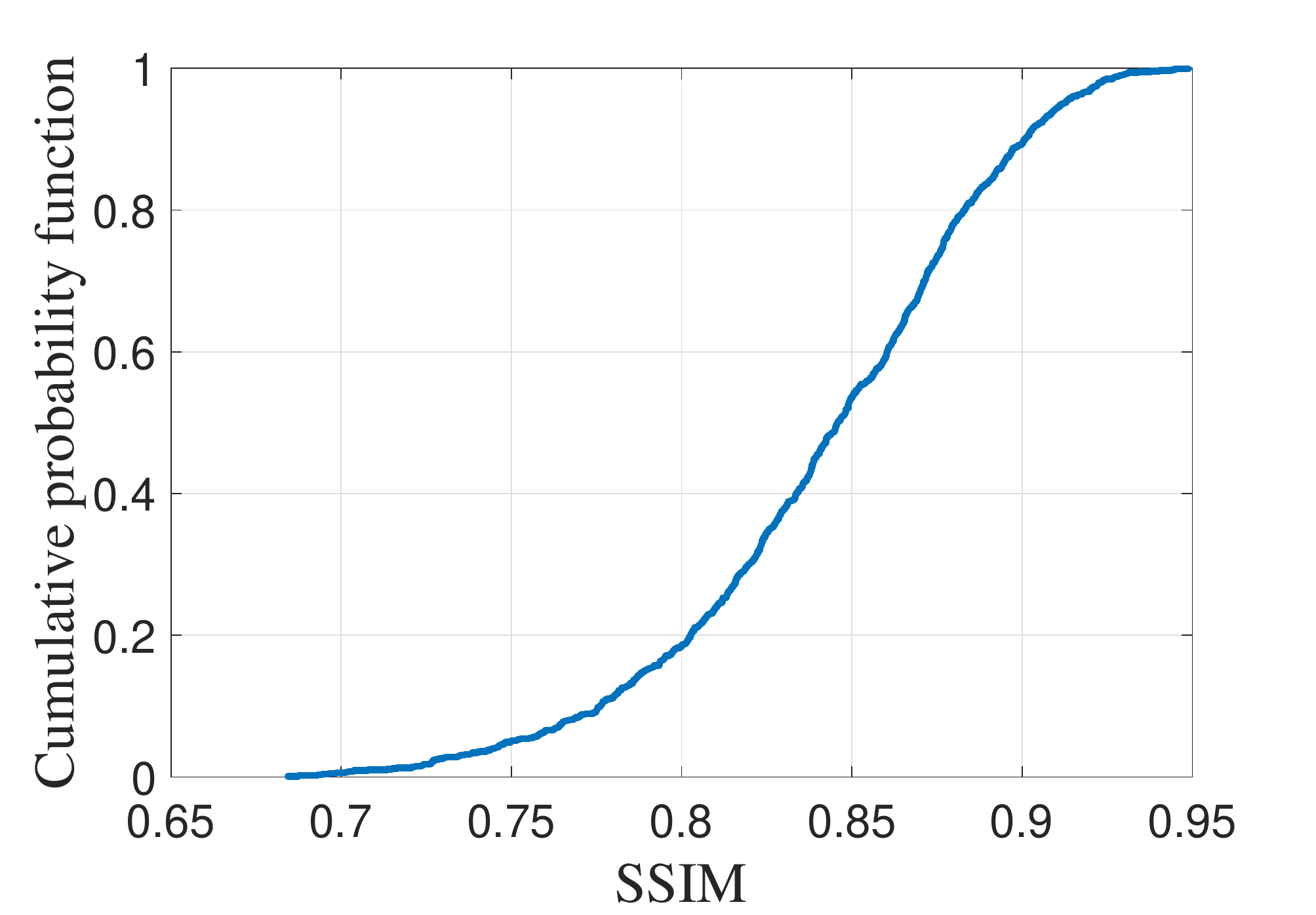}
    \caption{SSIM of PDAP for the generated channels.}
    \label{fig_ssim}
\end{figure}

\subsection{Angular Spread}
Angular spread describes how the power scatters in the spatial domain, which can be represented by
\begin{equation}
\begin{split}
\Bar{\theta}&=\frac{\sum_{i=0}^{N_{\theta}}i\Delta \theta P_{\theta}(i)}{\sum_{i=0}^{N_{\theta}}P_{\theta}(i)},\\
\theta_{rms}& = \sqrt{\frac{\sum_{i=0}^{N_{\theta}}(i\Delta \theta-\Bar{\theta})^2P_{\theta}(i)}{\sum_{i=0}^{N_{\theta}}P_{\theta}(i)}},
\end{split}
\end{equation}
where $N_{\theta}$ denotes the number of sampling points in the spatial domain, $\Bar{\theta}$ denotes the angle weighted by the power, $\theta_{rms}$ refers to the RMS angular spread, $\Delta \theta$ defines the angle interval, and $P_{\theta}(i)$ refers to the power at the AoA of $i\Delta\theta$. The CDF plot of angular spread for the original and generated channels is depicted in Fig.~\ref{fig_delay}(b). The CDF of angular spread for the generated channels has a good agreement with the original channels. Moreover, the mean values of angular spread for the original and generated channels are $56.82^{\circ}$ and $53.78^{\circ}$, respectively, which shows a small deviation. This suggests that the proposed T-GAN can well characterize the power distribution in the spatial domain.

\subsection{Power Delay Angular Profile}
Power delay angular profile characterizes the distribution of power in the spatial-temporal domain. In the experiment, the average power delay angular profiles (PDAPs) for the original and generated channels are compared as in Fig.~\ref{fig_pdap}. The generated channels shows good agreement with the original channels. To obtain a quantitative comparison, the deviation between the average PDAPs can be measured by root-mean-square error (RMSE), calculated as
\begin{equation}
\textrm{RMSE}=\sqrt{\frac{1}{N_{\tau}N_{\theta}}\sum(\overline{\textrm{PDAP}}_{r}(i,j)-\overline{\textrm{PDAP}}_{g}(i,j))^2},
\end{equation}
where $N_{\tau}$ and $N_{\theta}$ denote the number of sampling points in the temporal and spatial domains, respectively. Moreover, $\textrm{PDAP}(i,j)$ denotes the power at the delay of $i\Delta \tau$ and the AoA of $j\Delta \theta$, and 
 the terms $\overline{\textrm{PDAP}}_{r}$ and $\overline{\textrm{PDAP}}_{g}$ refer to the average PDAPs for the original and generated channels, respectively. The calculated RMSE value is 2.18 dB, which shows a small derivation, considering the wide range of the average PDAP values from -140 dB to -200 dB shown in Fig.~\ref{fig_pdap}. This proves that the proposed T-GAN can capture the features of channel in the joint spatial-temporal domain. This is attributed to the powerful capability of transformer structure in T-GAN to exploit the dependencies among different domains of channel.

Moreover, to measure the similarity quantitatively, Structure Similarity Index Measure (SSIM) is introduced, which is widely applied to evaluate the quality and similarity of images~\cite{ref_ssim}. The range of SSIM is from 0 to 1, and the value of SSIM is larger when the similarity between images is higher. The PDAPs of the generated channels are compared with the original channels at the same distance. The CDF of SSIM is shown in Fig.~\ref{fig_ssim}. It can be observed that the proposed T-GAN can achieve high SSIM values, which is above 0.8 for 80 percent of the generated channels. This further demonstrates the good performance of T-GAN in modeling the channels.

\section{Conclusion}\label{sec_conclusion}
In this paper, we proposed a T-GAN based THz spatial-temporal channel modeling method, which can capture the distribution of channel in the THz band. Moreover, the transformer structure is exploited in T-GAN to excavate dependencies among the channel parameters. Finally, we validate the performance of T-GAN with the THz dataset. T-GAN can generate the channels that have good agreement with the original channels in terms of delay spread, angular spread and power delay angular profile. With the capability of channel modeling and generating, T-GAN has the potential to assist the design of THz communication system, especially the end-to-end system.

\bibliographystyle{IEEEtran}
\bibliography{main}
\newpage

\end{document}